\documentclass[a4paper,12pt]{article}
\usepackage{graphicx,amssymb}

\newcommand{\bea}{\begin{eqnarray}}
\newcommand{\ena}{\end{eqnarray}}
\newcommand{\vs}[1]{\vspace{#1 mm}}
\newcommand{\hs}[1]{\hspace{#1 mm}}
\renewcommand{\a}{\alpha}
\renewcommand{\b}{\beta}
\renewcommand{\c}{\gamma}
\renewcommand{\d}{\delta}
\newcommand{\e}{\epsilon}
\newcommand{\s}{\sigma}

\def\bbox{{\,\lower0.9pt\vbox{\hrule \hbox{\vrule height 0.2 cm
\hskip 0.2 cm \vrule height 0.2 cm}\hrule}\,}}
\newcommand{\dsl}{\pa \kern-0.5em /}

\newcommand{\shalf}{\frac{1}{2}}
\newcommand{\pa}{\partial}
\newcommand{\nn}{\nonumber\\}
\newcommand{\p}[1]{(\ref{#1})}

\begin{document}

\renewcommand{\thefootnote}{\fnsymbol{footnote}}
\begin{titlepage}

\setcounter{page}{0}
\begin{flushright}
OU-HET 431 \\
hep-th/0302140
\end{flushright}

\vs{10}
\begin{center}
{\Large\bf Null-Brane Solutions in Supergravities}
\vs{30}

{\large
Nobuyoshi Ohta\footnote{e-mail address: ohta@phys.sci.osaka-u.ac.jp}}\\
\vs{10}
{\em Department of Physics, Osaka University,
Toyonaka, Osaka 560-0043, Japan}
\end{center}
\vs{15}
\centerline{{\bf{Abstract}}}
\vs{5}

We find a new class of time-dependent brane solutions in supergravities
in arbitrary dimensions $D$. These are general intersecting light-like
branes (null-branes), and their superposition and intersection rules
are obtained. This is achieved by directly solving bosonic
field equations for supergravity coupled to a dilaton and antisymmetric
tensor fields. We discuss their possible significance.

\end{titlepage}
\newpage
\renewcommand{\thefootnote}{\arabic{footnote}}
\setcounter{footnote}{0}
\setcounter{page}{2}

There has been much interest in time-dependent and spacelike brane solutions
(S-branes) of supergravities in eleven and ten dimensions because of
its possible connection with tachyon condensations and dS/CFT
correspondence~\cite{GS}-\cite{R} (see also refs.~\cite{LMP,H} for related
solutions). These theories are the low-energy limits of the string theories
and supposedly unifying M-theory of strings. Following the usual convention,
S$p$-branes are used for those with $(p+1)$-dimensional Euclidean world-volume.
The more general solutions can be understood as intersecting ones of these
fundamental S$p$-branes~\cite{DK,V,O} and the rules how the branes intersect
with each other are given in analogy to the usual branes~\cite{PT}--\cite{NO}.
Possible physical implications of these solutions are discussed in
\cite{CCK,BQR,GS2}.

The existence of S-brane solutions is inferred from the following
argument~\cite{GS}. Consider initial data at time $t=0$ with the tachyon
field sitting at the unstable maximum with a small velocity in a double-well
potential of unstable D3-brane in type IIA theory. As time evolves into
the future, the tachyon rolls off the top, emits closed string radiation
and settles down to the minimum. Evolving into the past, one finds the
time-reversed process with the tachyon approaching the other minimum.
The overall picture is that the finely tuned incoming radiation
conspires to excite the tachyon field to the top of the barrier and then
down to the other side, dissipating back into the radiation.
The obtained result is a timelike kink in the tachyon field.
This picture was used for constructing S-brane solutions in field
theories as well as superstrings/supergravities, and produces an interesting
class of time-dependent solutions. Time-dependent solutions are
investigated rather recently. It is thus interesting to try to find
other possible time-dependent $p$-brane solutions.

The above argument can be immediately used to argue the existence of
null-brane solutions (N-branes) with initial data on null
hypersurface~\cite{GS}. This class of solutions are also interesting
from the viewpoint of closed/open string correspondence and stringy
explanation of entropy of the black holes as discussed in ref.~\cite{KR},
where such solutions were discussed in the string worldsheet
picture. However, to the best of our knowledge,
no corresponding solutions in supergravities have been constructed.\footnote{
``N-brane'' solution is given in ref.~\cite{FS}, but it is called so
because the field strength has the component in null direction, and is
not the brane solution localized in the light-like direction.}
In order to gain insight into the geometric meaning, it is important to
have such solutions explicitly. It is the purpose of this note to give
this class of solutions in supergravities. In particular, we not only
construct such solutions but also give the intersection rules for the
way how the solutions can intersect with each other by extending the method
of \cite{O,NO}. We show that the rules are simple consequences of the
field equations, which can be easily integrated and the consistency of
the solutions reduces the problem of solving the field equations to
an algebraic one.

The results of our analysis turn out to be rather similar to
the superposition rules for other types of branes~\cite{O,NO}.
We show that the requirement that the field for each brane be
independent is sufficient to give the solutions and intersection rules.

Let us start with the general action for gravity coupled to a dilaton
$\phi$ and $m$ different $n_A$-form field strengths:
\bea
I = \frac{1}{16 \pi G_D} \int d^D x \sqrt{-g} \left[
 R - \shalf (\pa \phi)^2 - \sum_{A=1}^m \frac{1}{2 n_A!} e^{a_A \phi}
 F_{n_A}^2 \right].
\label{act}
\ena
This action describes the bosonic part of $D=11$ or $D=10$ supergravities;
we simply drop $\phi$ and put $a_A=0$ and $n_A=4$ for $D=11$, whereas we
set $a_A=-1$ for the NS-NS 3-form and $a_A=\shalf(5-n_A)$ for forms coming
from the R-R sector.\footnote{There may be Chern-Simons terms in the action,
but they are irrelevant in our following solutions.} To describe more
general supergravities in lower dimensions, we should include several scalars,
but for simplicity we disregard this complication in this paper.

{}From the action (\ref{act}), one derives the field equations
$$
R_{\mu\nu} = \shalf \pa_\mu \phi \pa_\nu \phi + \sum_{A} \frac{1}{2 n_A!}
 e^{a_A \phi} \left[ n_A \left( F_{n_A}^2 \right)_{\mu\nu}
 - \frac{n_A -1}{D-2} F_{n_A}^2 g_{\mu\nu} \right],
$$
$$
\bbox \phi = \sum_{A} \frac{a_A}{2 n_A!} e^{a_A \phi} F_{n_A}^2,
$$
$$
\pa_{\mu_1} \left( \sqrt{- g} e^{a_A \phi} F^{\mu_1 \cdots \mu_{n_A}} \right)
 = 0,
$$ \vs{-10}
\bea
\pa _{[\mu} F_{\mu_1 \cdots \mu_{n_A}]} = 0.
\label{fe}
\ena
The last equations are the Bianchi identities.

We take the following metric for our system:
\bea
ds_D^2 = -2 e^{2u_0} dudv - 2 e^{2u_1}dv^2 + \sum_{\a=2}^{p} e^{2 u_\a} dy_\a^2
 + e^{2B} d\Sigma_{k,\s}^2,
\label{met}
\ena
where $D=p+k+1$, the coordinates $v=(t+x)/\sqrt{2}$ and
$y_\a, (\a=2,\ldots, p)$ parametrize the $p$-dimensional world-volume
directions and the remaining coordinates of the $D$-dimensional spacetime
are the lightcone coordinate $u=(t-x)/\sqrt{2}$ and those for
$k$-dimensional spherical ($\s=+1$), flat ($\s=0$) or hyperbolic
($\s=-1$) spaces, whose line elements are $d\Sigma_{k,\s}^2$.
Since we are interested in solutions localized in the lightcone directions,
all the functions appearing in the metrics as well as dilaton $\phi$ are
assumed to depend only on $u$.
The Ricci tensors for the metric~\p{met} are
\bea
R_{uu} &=& - \sum_{\a=2}^p [ u_\a'' + (u_\a')^2 - 2u_\a' u_0']
 - k[ B'' + (B')^2 - 2 B' u_0'], \nn
R_{vv} &=& 4 e^{-4u_0 +4 u_1} \Big[ u_1'' + u_1'(-2u_0' +2u_1'
 + \sum_{\a=2}^p u_\a' + k B' )\Big], \nn
R_{uv} &=& 2 e^{-2u_0 +2 u_1} \Big[ u_1'' + u_1'(-2u_0' +2u_1'
 + \sum_{\a=2}^p u_\a' + k B' )\Big], \nn
R_{\a\b} &=& -2 e^{2(-2u_0 +u_1+ u_\a)} \Big[ u_\a'' + u_\a'( -2 u_0'
 +2u_1' + \sum_{\c=2}^p u_\c' + k B') \Big]\d_{\a\b}, \nn
R_{ab} &=& -2 e^{2(-2u_0+u_1+B)}\Big[ B'' + B'( -2u_0' + 2u_1'
 + \sum_{\a=2}^p u_\a' +k B')\Big]\bar g_{ab} + \s (k-1)\bar g_{ab},
\label{ricci}
\ena
where $\bar g_{ab}$ is the metric for the hypersurface $\Sigma_{k,\s}$.
Here and in what follows, a prime denotes a derivative with
respect to $u$. We note that the above metric~\p{met} is similar to but
not quite the same as what is considered in the pp-wave solutions~\cite{pp}
with the metric $ds_D^2 = -2 e^{2u_0} dudv - 2 e^{2u_1}du^2 + \cdots$.
One may wonder if any interesting solutions exist for this case or for the
metric~\p{met} without $dv^2$ term. We find that neither of these metrics
give solutions with nontrivial field strengths corresponding to N-branes
of our interest.

For the field strengths, we take the most general ones consistent
with the field equations and Bianchi identities.
Those for an electrically charged N$q$-brane (whose world-volume
is $(q+1)$-dimensional) is given by
\bea
F_{uv \a_2 \cdots \a_{q+1}} = \e_{v\a_2 \cdots \a_{q+1}} E', \hs{3}
(n_A = q+2),
\label{ele}
\ena
where $v, \a_2, \cdots ,\a_{q+1}$ stand for the tangential directions to
the N$q$-brane. The magnetic case is given by
\bea
F^{\a_{q+2} \cdots \a_p a_1 \cdots a_{k}} = \frac{1}{\sqrt{-g}}
 e^{-a\phi} \e^{\a_{q+2} \cdots \a_p a_1 \cdots a_k} {\tilde E}',
\hs{3} (n_A = D-q-2)
\label{mag}
\ena
where $a_1, \cdots, a_{k}$ denote the coordinates of the $k$-dimensional
hypersurface $\Sigma_{k,\s}$.
The functions $E$ and $\tilde E$ are again assumed to depend only on $u$.

The electric field (\ref{ele}) trivially satisfies the Bianchi
identities but the field equations are nontrivial. On the other hand, the
field equations are trivial but the Bianchi identities are nontrivial
for the magnetic field (\ref{mag}).

We will solve the field eqs.~(\ref{fe}) with the ansatz
\bea
2 u_0 = 2u_1+ \sum_{\a=2}^p u_\a + k B,
\label{ans}
\ena
which simplifies the field equations~(\ref{fe}) considerably.
For both cases of electric~(\ref{ele}) and magnetic~(\ref{mag})
fields, we find that the field eqs.~(\ref{fe}) are cast into
\bea
&& - 2u_0'' + 2u_1'' + 4u_0'(u_0'-u_1') - \sum_{\a=2}^p (u_\a')^2 - k(B')^2
= \frac{1}{2} \phi'^2,
\label{1}
\\
&&
u_1'' = \sum_{A} \frac{D-q_A-3}{4(D-2)} S_A (E_A')^2,
\label{2'}
\\
&&
u_\a'' = \sum_{A} \frac{\d_A^{(\a)}}{4(D-2)} S_A (E_A')^2,
 \hs{3} (\a=2,\cdots,p),
\label{2}
\\
&& B'' - \frac{\s (k-1)}{2} e^{4u_0-2u_1-2B} = -\sum_{A} \frac{q_A+1}{4(D-2)}
 S_A (E_A')^2,
\label{3}
\\
&& \phi'' = -\sum_{A} \frac{\e_A a_A}{4} S_A (E_A')^2,
\label{4}
\\
&& \left( S_A E_A' \right)' = 0,
\label{5}
\ena
where $A$ denotes the kinds of $q_A$-branes and we have defined
\bea
S_A \equiv \exp \left( \e_A a_A \phi - 2 \sum_{\a \in q_A} u_\a \right),
\label{6}
\ena
(here and in what follows $\a=1$ is included in the sum when written as
$\a\in q_A$) and
\bea
\d_A^{(\a)} = \left\{ \begin{array}{l}
D - q_A - 3 \\
- (q_A+1)
\end{array}
\right.
\hs{5}
{\rm for} \hs{3}
\left\{
\begin{array}{l}
y_\a \mbox{  belonging to $q_A$-brane} \\
{\rm otherwise}
\end{array},
\right.
\ena
and $\e_A= +1 (-1)$ corresponds to electric (magnetic) fields.
For magnetic case we have dropped the tilde from $E_A$. Equations~\p{1},
\p{2'}, \p{2} and \p{3} are the $uu, vv, \a\a$ and $ab$ components of the
Einstein equation in \p{fe}, respectively. ($uv$ and $vv$ components give
the same eq.~\p{2'}). We also define $\d_A^{(1)} = D - q_A - 3$, so that
eq.~\p{2'} can be written as \p{2}. The last one is the field equation for
the field strengths of the electric fields and/or Bianchi identity
for the magnetic ones. It is remarkable that both the electric and magnetic
cases can be treated simultaneously just by using the sign $\e_A$. This is
because the original system~\p{act} has the S-duality symmetry under
\bea
g_{\mu\nu} \to g_{\mu\nu}, \quad
F_{n_A} \to e^{-a_A \phi}*\! F_{n_A}, \quad
\phi \to - \phi.
\label{sdual}
\ena

{}From eq.~(\ref{5}) one finds
\bea
S_A E_A' = c_A,
\label{const}
\ena
where $c_A$ is a constant.
With the help of eq.~\p{const}, we find that eqs.~\p{2} and \p{4} give
\bea
u_\a' &=& \sum_{A} \frac{\d_{A}^{(\a)}}{4(D-2)} c_A E_A + c_\a, \quad
(\a=1, \cdots, p), \nn
\phi' &=& - \sum_{A} \frac{\e_A a_A}{4} c_A E_A + c_\phi,
\label{fint}
\ena
where $c_\a$ and $c_\phi$ are integration constants.
Let us next define
\bea
g(u) = (2u_0-u_1-B)/(k-1).
\label{gdef}
\ena
We find from \p{ans}
\bea
B = g - \frac{1}{k-1}\sum_{\a=1}^p u_\a, \qquad
2u_0 = kg +u_1 - \frac{1}{k-1}\sum_{\a=1}^p u_\a,
\ena
Using \p{fint}, we get
\bea
\label{bdot}
B' &=& g' - \sum_{A} \frac{q_A+1}{4(D-2)} c_A E_A
 - \frac{1}{k-1}\sum_{\a=1}^p c_\a, \\
2u_0' &=& k g' + \sum_{A} \frac{D-2q_A-4}{4(D-2)} c_A E_A
 + c_1 - \frac{1}{k-1}\sum_{\a=1}^p c_\a,
\label{u0dot}
\ena
Substituting \p{const}, \p{gdef} and \p{bdot} into \p{3}, we obtain
\bea
g'' - \frac{\s(k-1)}{2} e^{2(k-1)g} = 0,
\label{gddot}
\ena
which yields
\bea
g'^2 -\frac{\s}{2} e^{2(k-1)g} = \b^2,
\label{gdot}
\ena
where $\b$ is an integration constant. The solution to eq.~\p{gdot} is
given by
\bea
g(u) = \left\{\begin{array}{ll}
\frac{1}{k-1} \ln \frac{\sqrt{2}\b}{\sinh[(k-1)\b(u-u_1)]} & :\s=+1, \\
\pm \b(u-u_1) & :\s=0, \\
\frac{1}{k-1} \ln \frac{\sqrt{2}\b}{\cosh[(k-1)\b(u-u_1)]} & :\s=-1,
\end{array}
\right.
\label{g}
\ena
where $u_1$ is another integration constant.

Substituting eqs.~(\ref{fint}) and \p{bdot}-\p{gdot} into
(\ref{1}) yields
\bea
&& \left( \sum_{A} \frac{D-2q_A-4}{4(D-2)} c_A E_A
+ c_1 -\frac{1}{k-1}\sum_{\a=1}^p c_\a\right) \left( \sum_{A} \frac{c_A E_A}{4}
 + c_1+\frac{1}{k-1}\sum_{\a=1}^p c_\a \right) \nn
&& + \; \sum_{\a=2}^p \left( \sum_{A} \frac{\d_{A}^{(\a)}}{4(D-2)} c_A E_A
 + c_\a \right)^2 + k \left( \sum_{A} \frac{q_A+1}{4(D-2)} c_A E_A
 + \frac{1}{k-1}\sum_{\a=1}^p c_\a\right)^2 \nn
&& + \; \shalf \left( \sum_{A} \frac{\e_A a_A}{4} c_A E_A - c_\phi \right)^2
 - \sum_{A} \frac{c_A}{4} E_A' - k(k-1) \b^2 =0.
\label{sint}
\ena
This equation must be valid for arbitrary functions $E_A$ of $u$.
{}From the $E_A$-independent part of eq.~(\ref{sint}), one finds
\bea
\frac{1}{k-1}\left(\sum_{\a=1}^p c_\a\right)^2
 + \sum_{\a=1}^p c_\a^2 + \shalf c_\phi^2= k(k-1) \b^2.
\label{condconst}
\ena
We can then rewrite eq.~(\ref{sint}) as
\bea
\sum_{A,B} \left[ M_{AB} \frac{c_A}{4} + \d_{AB} \left\{
\left( \frac{1}{E_A}\right)' + \frac{2\tilde c_A}{E_A}\right\} \right]
c_B E_A E_B =0,
\label{tint}
\ena
where
\bea
\label{cond1}
M_{AB} &=& \frac{D-q_A-q_B-4}{D-2}+\sum_{\a=2}^p
\frac{\d_{A}^{(\a)}\d_{B}^{(\a)}}{(D-2)^2}
 + k \frac{(q_A+1)(q_B+1)}{(D-2)^2} + \shalf \e_A a_A \e_B a_B, \\
\tilde c_A &=& \sum_{\a\in q_A} c_\a-\frac{1}{2} c_\phi \e_A a_A.
\label{cond11}
\ena
Since $M_{AB}$ is constant, eq.~(\ref{tint}) cannot be satisfied for
arbitrary functions $E_A$ of $u$ unless the second term inside
the square bracket is a constant. Requiring this to be a constant tells
us that the function $E_A$ must satisfy
\bea
\left(\frac{1}{E_A}\right)' + \frac{2\tilde c_A}{E_A} + \tilde c_A N_A=0,
\label{har}
\ena
or
\bea
E_A = - \frac{e^{\tilde c_A(u-u_A)}}{N_A \cosh \tilde c_A(u-u_A)},
\label{har1}
\ena
where $N_A$ is a normalization factor and $u_A$ is an integration constant.
In this way, the problem reduces to the algebraic equation (\ref{tint})
supplemented by (\ref{har}) without making any assumption other than
(\ref{ans}).

Equation~(\ref{tint}) has two implications if we take
independent functions for the fields $E_A$. In this case,
first putting $A=B$ in eq.~(\ref{tint}), we learn that
\bea
c_A = \frac{4(D-2)\tilde c_A N_A}{\Delta_A},
\label{cond2}
\ena
where
\bea
\Delta_A = (q_A + 1) (D-q_A-3) + \shalf a_A^2 (D-2).
\label{res2}
\ena
By use of eqs.~\p{har1} and \p{cond2}, eqs.~\p{fint}, \p{bdot} and \p{u0dot}
can be integrated with the results
\bea
2u_0 &=& kg(u) - \sum_{A} \frac{D-2q_A-4}{\Delta_A} \ln \cosh\tilde c_A(u-u_A)
 - c_0 u + c_0', \nn
u_\a &=& - \sum_{A} \frac{\d_{A}^{(\a)}}{\Delta_A} \ln \cosh\tilde c_A(u-u_A)
 - \tilde c_\a u + c_\a', \hs{3} (\a=1,\cdots, p), \nn
B &=& g(u) + \sum_{A} \frac{q_A+1}{\Delta_A} \ln \cosh\tilde c_A(u-u_A)
 + c_b u +c_b', \nn
\phi &=& \sum_{A} \frac{(D-2)\e_A a_A}{\Delta_A} \ln \cosh\tilde c_A(u-u_A)
 + \tilde c_\phi u + c_\phi',
\label{res3}
\ena
where $c'$'s are new integration constants and
\bea
&& c_0 = \sum_A \frac{D-2q_A-4}{\Delta_A}\tilde c_A -c_1
 + \frac{\sum_{\a=1}^p c_\a}{k-1}, \;\;
c_0' =c_1' - \frac{\sum_{\a=1}^p c_\a'}{k-1}, \;\;
\tilde c_\a = \sum_A \frac{\d_{A}^{(\a)}}{\Delta_A}\tilde c_A-c_\a, \nn
&& c_b=\sum_A \frac{q_A+1}{\Delta_A}\tilde c_A-\frac{\sum_{\a=1}^p c_\a}{k-1},
\;\; c_b' = -\frac{\sum_{\a=1}^p c_\a'}{k-1},\;\;
\tilde c_\phi = \sum_A \frac{(D-2)\e_A a_A}{\Delta_A}\tilde c_A +c_\phi.
\ena

To fix the normalization $N_A$, we go back to eq.~(\ref{6}).
Using (\ref{res3}), we find
\bea
S_A = [\cosh\tilde c_A(u-u_A)]^2 e^{\e_Aa_A c_\phi'-2\sum_{\a\in q_A} c_\a'},
\label{ress}
\ena
which, together with (\ref{const}) and (\ref{cond2}), leads to
\bea
N_A = \sqrt{\frac{\Delta_A}{4(D-2)}}
e^{\e_Aa_A c_\phi'/2-\sum_{\a\in q_A} c_\a'}.
\label{norm}
\ena

Our metric and other fields are thus finally given by
\bea
ds_D^2 &=& \prod_A [\cosh\tilde c_A (u-u_A)]^{2 \frac{q_A+1}{\Delta_A}}
 \Bigg[ - 2 e^{kg(u)-2c_0 u+2c_0'}
 \prod_A [\cosh\tilde c_A (u-u_A)]^{-\frac{D-2}{\Delta_A}} dudv \nn
&& \hs{10} - \; 2 \prod_A [\cosh\tilde c_A (u-u_A)]^{-2\frac{D-2}{\Delta_A}}
 e^{-2\tilde c_1 u+2c_1'}dv^2 + e^{2g(u)+2c_bu+2c_b'}d\Sigma_{k,\s}^2 \nn
&& \hs{10} + \; \sum_{\a=2}^{p} \prod_A [\cosh\tilde c_A(u-u_A)]^{- 2
 \frac{\c_A^{(\a)}}{\Delta_A}} e^{-2 \tilde c_\a u+2c_\a'} dy_\a^2\Bigg], \nn
E_A &=& -\frac{e^{\tilde c_A(u-u_A)}}{N_A \cosh \tilde c_A(u-u_A)},\quad
 \tilde c_A = \sum_{\a\in q_A} c_\a-\frac{1}{2} c_\phi \e_A a_A.
\ena
where we have defined
\bea
\c_A^{(\a)} = \left\{ \begin{array}{l}
D-2 \\
0
\end{array}
\right.
\hs{5}
{\rm for} \hs{3}
\left\{
\begin{array}{l}
y_\a \hs{3} {\rm belonging \hs{2} to} \hs{2} q_A{\rm -brane} \\
{\rm otherwise}
\end{array}.
\right.
\ena
These solutions contain $2p+2$ integration constants $c_\a, c_\a'
(\a=1, \cdots,p), c_\phi, c_\phi'$, together with $u_1$ and $u_A$ with $\b$
determined by eq.~\p{condconst}. Among these, $c_\a'$ can be removed by
rescaling the coordinates, and $u_1$ by a shift of the coordinate $u$.
Without any preference of the choice of other parameters, we leave
these as free parameters. Thus the general solutions can be constructed by
the following rules:\\
(1) All the directions are multiplied by $[\cosh\tilde c_A(u-u_A)]^{2
\frac{q_A+1}{\Delta_A}}$, and in addition,\\
(2) the overall transverse directions ($u$ and $k$-dimensional space) are
multiplied by $e^{kg(u)}$ and $e^{2g(u)}$, respectively,\\
(3) the coordinates belonging to the brane are multiplied by
$[\cosh \tilde c_A(u-u_A)]^{-\frac{D-2}{\Delta_A}}$.

The second condition following from eqs.~(\ref{tint}) is $M_{AB}=0$ for
$A \neq B$. This leads to the intersection rules for two branes:
If $q_A$-brane and $q_B$-brane intersect over ${\bar q} (\leq q_A, q_B)$
dimensions, this gives
\bea
{\bar q} = \frac{(q_A+1)(q_B+1)}{D-2}-1 - \shalf \e_A a_A \e_B a_B.
\label{int}
\ena
Remember that the world-volume of $q$-branes lies in $(q+1)$-dimensional
space including $v$. For eleven-dimensional supergravity, we have electric
N2-branes, magnetic N5-branes and no dilaton $a_A=0$. The rule \p{int}
tells us that N2-brane can intersect with N2-brane over a `0-brane'
$(\bar q = 0)$ (which actually lives in 1-dimensional space $v$) and with
N5-brane over a `string' $(\bar q=1)$ (2-dimensional space including $v$),
and N5-brane can intersect with N5-brane over `3-brane' $(\bar q=3)$
(4-dimensional space). In particular, our results show that
there is no other intersecting solution as long as we treat the functions
$E_A$ with different index $A$ as independent. If this condition is relaxed,
there may be other solutions. This is again quite similar to the
intersection rules for usual branes~\cite{PT}-\cite{NO} and S-branes~\cite{O}.

For all the light-like D$q$-brane solutions in type II superstrings, we find
\bea
\e a = \frac{3-q}{2},
\ena
which tells us that the intersection rule is
\bea
\bar q = \frac{q_A+q_B}{2}-2.
\ena

It may be instructive to see how a single N-brane solution looks like.
The metrics for the N2- and N5-branes in $D=11$ supergravity take the form
\bea
ds_{N2}^2 &=& [\cosh\tilde c (u-u_2)]^{1/3}
 \Bigg[ -2 e^{7g(u)+2c u} [\cosh\tilde c (u-u_2)]^{-1/2} dudv \nn
&& +\; [\cosh\tilde c (u-u_2)]^{-1} \{-2 e^{4c u} dv^2
 + e^{-2 c u}(dy_2^2 +dy_3^2)\}
 + e^{2g(u)+2c'}d\Sigma_{7,\s}^2 \Bigg], \nn
&& \tilde c^2 + 12 c^2 = 84 \b^2, \nn
ds_{N5}^2 &=& [\cosh\tilde c (u-u_2)]^{2/3}
 \Bigg[ -2 e^{4g(u)+5c u} [\cosh\tilde c (u-u_2)]^{-1/2} dudv \nn
&& \hs{-5} + \; [\cosh\tilde c (u-u_2)]^{-1} \{-2 e^{10c u} dv^2
+ e^{-2 c u}(dy_2^2 +\cdots + dy_6^2)\}+ e^{2g(u)+2c'}d\Sigma_{4,\s}^2 \Bigg],
\nn
&& \tilde c^2 + 60 c^2 = 24 \b^2,
\ena
where we have imposed the rotational symmetry on the brane world-volume
$y_\a$, absorbed some constants in the metric by the rescaling of the
coordinates and redefined parameters. With eq.~\p{g}, these solutions have
5 independent parameters $\tilde c,c,c',u_1$ and $u_2$, but $u_1$ may be
eliminated by the shift of the coordinate $u$, resulting in 4 parameters.
A more interesting solution is the ND3-brane in type IIB:
\bea
ds_{ND3}^2 &=& -2 e^{5g(u)+3c u} dudv + [\cosh\tilde c (u-u_2)]^{-1/2}
 \{-2 e^{6c u}dv^2 + e^{-2c u}(dy_2^2 + \cdots + dy_4^2)\} \nn
&& + \; e^{2g(u)+2c'}[\cosh\tilde c (u-u_2)]^{1/2} d\Sigma_{5,\s}^2,\qquad
\tilde c^2 + 24 c^2 +c_\phi^2 = 40 \b^2,
\ena
where again rotational symmetry on the world-volume is imposed.

We have examined if these solutions preserve any supersymmetry. It turns out
that there is no remaining supersymmetry in these solutions, similarly to
S-branes. In fact, they are supposed to correspond to branes with Dirichlet
boundary conditions in the lightcone direction, and hence describe
configurations which exist only for a fixed lightcone coordinate.
It would be interesting to examine stability and particle creations
in these geometry~\cite{BQR}.

To summarize, we have given quite a general model-independent derivation
of the N-brane solutions in supergravities in arbitrary dimensions.
The intersection rules simply follow from the field equations if we require
that the functions $E_A$ with different index $A$ be independent.
In all cases, the algebraic eq.~(\ref{tint}) (together with (\ref{har}))
must be satisfied, and this equation should be most useful to examine
possible solutions. Our derivation is a simple generalization of the
general method developed in refs.~\cite{NO,O}. It is quite satisfying to
see that this is such a useful method.
We hope to discuss various properties of these solutions using the hints
from dualities implied by underlying string dynamics elsewhere.

\section*{Acknowledgement}

This work was supported in part by Grants-in-Aid for Scientific Research
Nos. 12640270 and 02041.

\newcommand{\NP}[1]{Nucl.\ Phys.\ B\ {\bf #1}}
\newcommand{\PL}[1]{Phys.\ Lett.\ B\ {\bf #1}}
\newcommand{\CQG}[1]{Class.\ Quant.\ Grav.\ {\bf #1}}
\newcommand{\CMP}[1]{Comm.\ Math.\ Phys.\ {\bf #1}}
\newcommand{\IJMP}[1]{Int.\ Jour.\ Mod.\ Phys.\ {\bf #1}}
\newcommand{\JHEP}[1]{JHEP\ {\bf #1}}
\newcommand{\PR}[1]{Phys.\ Rev.\ D\ {\bf #1}}
\newcommand{\PRL}[1]{Phys.\ Rev.\ Lett.\ {\bf #1}}
\newcommand{\PRE}[1]{Phys.\ Rep.\ {\bf #1}}
\newcommand{\PTP}[1]{Prog.\ Theor.\ Phys.\ {\bf #1}}
\newcommand{\PTPS}[1]{Prog.\ Theor.\ Phys.\ Suppl.\ {\bf #1}}
\newcommand{\MPL}[1]{Mod.\ Phys.\ Lett.\ {\bf #1}}
\newcommand{\JP}[1]{Jour.\ Phys.\ {\bf #1}}

\end{document}